# Exchange Biasing of the Ferromagnetic Semiconductor $Ga_{1-x}Mn_xAs$

K. F. Eid, M. B. Stone, K. C. Ku, P. Schiffer and N. Samarth*

*Department of Physics and Materials Research Institute, Pennsylvania State University, University Park, PA 16802*

T. C. Shih and C. J. Palmstrøm

*Department of Chemical Engineering and Materials Science, University of Minnesota, Minneapolis MN 55455*

## ABSTRACT

We demonstrate the exchange coupling of a ferromagnetic semiconductor ($Ga_{1-x}Mn_xAs$) with an overgrown antiferromagnet (MnO). Unlike most conventional exchange biased systems, the blocking temperature of the antiferromagnet ($T_B = 48 \pm 2$ K) and the Curie temperature of the ferromagnet ($T_C = 55.1 \pm 0.2$ K) are comparable. The resulting exchange bias manifests itself as a clear shift in the magnetization hysteresis loop when the bilayer is cooled in the presence of an applied magnetic field and an enhancement of the coercive field.

* nsamarth@phys.psu.edu

Ferromagnetic semiconductors (FMSs) are the focus of extensive research for applications in semiconductor spintronics [1,2]. These hybrid materials offer new functionality because their carrier-mediated ferromagnetism can be manipulated by employing established techniques used in semiconductor technology [3]. Several device geometries have been demonstrated using the "canonical" FMS $Ga_{1-x}Mn_xAs$, including magnetic tunnel junctions [4,5,6], spin LEDs [7] and spin-tunable resonant tunnel diodes [8]. It is important in this context to develop a means of "exchange biasing" ferromagnetic semiconductors for device applications. This entails exchange coupling the FMS to a proximal antiferromagnet (AFM), as has been observed in ferromagnetic (FM) metals, intermetallic alloys, and oxides [9,10,11,12].

The hallmark of an exchange biased system is a hysteresis loop which is not centered about zero magnetic field, but is shifted either to positive or negative fields depending upon the orientation of the applied field while cooling the sample through the blocking temperature ($T_B$) of the AFM layer. This is typically accompanied by an increase in the coercive field ($H_C$) of the FM. In most examples of exchange bias, $T_B \leq T_N \ll T_C$ where $T_N$ is the Néel temperature of the AFM and $T_C$ is the Curie temperature of the FM. Ferromagnetic semiconductors such as $Ga_{1-x}Mn_xAs$ are of interest because $T_C$ can be varied from ~10 K to ~150 K by varying parameters such as Mn concentration and sample thickness [13]; in principle, this allows studies of exchange bias in a FM where $T_B$ can be above or below $T_C$.

We are aware of only one reported attempt to exchange bias a FMS, using MnTe as the proximal AFM on top of $Ga_{1-x}Mn_xAs$ [14]. Although the FMS showed an increased $H_C$, there was no conclusive evidence for an exchange coupling between the FMS and the AFM. We demonstrate here unambiguous exchange bias in a $Ga_{1-x}Mn_xAs$ epilayer on which a thin layer of Mn has been deposited. Rutherford backscattering scattering (RBS) measurements show that -- upon removal from the vacuum chamber -- the layer of Mn oxidizes to form MnO (an AFM with $T_N$ = 118 K [15]). While our experiments show clear evidence that $Ga_{1-x}Mn_xAs$ can indeed be exchange-biased by MnO, we also find that the high reactivity between Mn and GaAs requires better control over the interfacial characteristics between the two materials.

Samples are grown by low temperature molecular beam epitaxy on (001) semi-insulating, epi-ready GaAs substrates bonded to a mounting block by indium. After thermal desorption of the oxide layer, a high temperature GaAs layer is grown under standard conditions, followed by a low temperature GaAs layer grown at 250 °C. The magnetically active region of the sample consists of a 10 nm thick $Ga_{1-x}Mn_xAs$ layer ($x$ = 0.08), capped with a thin (~ 4 nm) Mn layer. After the growth of the $Ga_{1-x}Mn_xAs$ layer, the sample is moved from the growth chamber into an adjoining high vacuum buffer chamber and the As cell is cooled to room temperature. Once the As pressure in the growth chamber decreases to an acceptable level, the substrate is reintroduced into the growth chamber, and the Mn capping layer is grown at room temperature. Alternatively, the sample is transferred in vacuum to an adjacent As-free growth chamber for the Mn capping. Exchange-biased samples have been obtained using both schemes. These precautions are to avoid the formation of FM MnAs during the growth of the capping layer.

The epitaxial growth is monitored by *in-situ* reflection high energy electron diffraction (RHEED) at 12 keV. The $Ga_{1-x}Mn_xAs$ surface shows a clear reconstructed (1 x 2) RHEED pattern; the Mn surface shows a weaker, yet streaky and unreconstructed pattern whose symmetry is suggestive of the stabilization of a cubic phase of Mn. The thickness of the $Ga_{1-x}Mn_xAs$ layer is calculated from RHEED oscillations, while the thickness of the Mn overlayer is estimated from RHEED oscillations of MnAs whose growth rate is mainly determined by the sticking coefficient of Mn. The Mn concentration in the ferromagnetic layer is estimated at $x \sim 0.08$, based upon a calibration as a function of the Mn cell temperature ($T = 785\ ^oC$) [16]. Magnetization measurements are performed using a SQUID magnetometer, where samples are either field cooled or zero field cooled from room temperature to the measuring temperature in a finite or zero magnetic field respectively.

The post-growth removal of the samples from the mounting block requires a modest annealing cycle for a few minutes at $\sim 200\ ^0C$: the implications of this will be discussed later. The top Mn layer oxidizes upon exposure to air forming MnO. We have observed little change in the magnetization measurements of exchange biased samples over the course of six months, suggesting that the entire Mn layer was already oxidized at the time of the first measurement. The nature of the Mn layer is probed by RBS analysis using 2.3 MeV and 1.4 MeV $He^+$ ions with both normal and glancing angle detector geometries, corresponding to scattering angles of 165° and 108°, respectively. Both random and <001> channeling measurements are conducted. The channeling measurements show that the Mn overlayer is completely oxidized with stoichiometry close to $MnO_x$ (x ~1 to 1.5); the thickness of this MnO layer is consistent with the estimate from RHEED oscillations.

Figures 1(a) and 1(b) depict two hysteresis loops for the $Ga_{1-x}Mn_xAs$/MnO bilayer, measured at $T$ = 10 K after cooling the sample in a magnetic field of $H$ = ±2500 Oe. The "horizontal" offset of the hysteresis loops from the origin in a direction opposite to that of the cooling field is a clear signature of exchange coupling in the bilayer. Control measurements are used to rule out extrinsic effects: Fig. 1 (c) shows that cooling the sample in zero applied magnetic field results in a slightly biased hysteresis loop. The presence of some bias suggests that a small remnant magnetic field was present in the SQUID while cooling the sample. Such a field partially magnetizes the FM layer, which in turn sets the bias in the AFM layer. The sheared appearance of the magnetization curve suggests that there are multiple domains in the FM and that the sample was very close to being zero-field and zero-magnetization cooled. Fig. 1 (d) shows that the exchange bias shift is absent in a sample of similar thickness and grown under identical conditions, but without the AFM overlayer. There is also a notable increase of the coercive field for the Mn-capped sample compared to typical values for uncapped samples [17], which is consistent with the expected effects of exchange biasing [9].

Figures 2(a) and (b) compare the field-cooled hysteresis loops at 5 K and 30 K, respectively. In these plots, the sample is cooled from $T > T_C$ to the measuring temperature in a field of $H$ = 2500 Oe. With increasing temperature, the hysteresis loops become narrower, and their centers move closer to the origin. As the temperature approaches $T_C$, the hysteresis loop collapses into a single non-hysteretic curve of zero coercivity (not shown), indicating that the $Ga_{1-x}Mn_xAs$ layer has become paramagnetic.

We do not observe any training effects for a hysteresis loop performed multiple times at a particular temperature after a single field cooling of the sample.

Figure 2(c) summarizes the temperature dependence of the hysteresis loop parameters $H_E$ and $H_C$. From this figure, we estimate $T_B = 48 \pm 2$ K. The temperature dependence of the saturation magnetization (Fig. 2(d)) indicates that $T_C = 55.1 \pm 0.2$ K; both the value of $T_C$ and the magnitude of the saturation magnetization are similar for the uncapped and capped samples, indicating that the Mn overlayer does not affect the FM state of the $Ga_{1-x}Mn_xAs$ layer except through the exchange biasing. Fig. 2(c) shows that $H_E$ undergoes a monotonic decrease with temperature until vanishing at $T_B$, while $H_C$ passes through a broad peak near $T_B$ before becoming zero at $T_C$. The absence of a second peak in $H_E$ at $T_C$ and of any exchange bias in the paramagnetic state further suggest that $T_C > T_B$. [18].

Figure 3 shows the dependence of $H_E$ and $H_C$ on the magnitude of the cooling field. Cooling fields as small as $H = 25$ Oe produce a significant exchange bias, and both $H_E$ and $H_C$ vary only slightly with the magnitude of cooling field up to $H \gg H_E, H_C$. These data indicate that once exchange bias has been established, it becomes insensitive to the magnitude of the cooling field. Establishing exchange bias requires only a field large enough to saturate the ferromagnetic layer at temperatures above $T_B$. Another important observation is that although the coupling energy in this system ($\sim 3\times10^{-3}$ erg/cm$^2$) is small compared to that in typical exchange bias systems, we observe relatively large exchange bias fields. This is because the magnetization of the FM layer is relatively low.

Finally, we comment on the reproducibility of our results: the data shown in the figures are from a single sample, and we have observed exchange biasing in one other sample out of the ten that were grown under similar conditions. We attribute the difficulty in reproducing the results to the extreme reactivity of Mn with GaAs [19]. This could result in the formation of a detrimental interfacial layer when the samples are removed from the mounting block. To test this hypothesis, we intentionally annealed a sample that shows exchange bias for 10 minutes at 220 $^0$C; although these conditions are only slightly more severe than those typically used for sample removal from the mounting block, they are sufficient to completely eliminate the exchange bias. RBS analysis on this annealed sample still shows a MnO overlayer with similar characteristics to the exchange biased sample; however, we find a slight increase in the channeling minimum yield and also a small enhancement of the Ga channeling interface peak by ~1x10$^{15}$ atoms/cm$^2$. This suggests the formation of a thin Mn-GaAs interfacial reacted layer, consistent with studies of thermally induced reactions between Mn and GaAs [19]. Since many other choices for the AFM exchange biasing layer also involve Mn (such as FeMn, NiMn, MnSe, and MnTe), it is likely that interfacial reactivity will play an important role in other explorations of exchange biasing $Ga_{1-x}Mn_xAs$. For instance, we have thus far not seen any credible indications of either exchange coupling or exchange biasing of $Ga_{1-x}Mn_xAs$ using MnSe. In addition, we have attempted growths of Mn-capped $Ga_{1-x}Mn_xAs$ followed by an overgrown layer of Al to prevent oxidation; these experiments also have not yielded any measurable exchange bias. Nonetheless, the demonstrated feasibility of exchange biasing of this FMS -- as well as the identification of the accompanying difficulties – provides an important starting point for additional experiments.

This research has been supported by the DARPA-SPINS program under grant numbers N00014-99-1093, -99-1-1005, -00-1-0951, and -01-1-0830, by ONR N00014-99-1-0071 and by NSF DMR 01-01318. We thank J. Bass for useful discussions.

**Figure 1.** Hysteresis loops of a $Ga_{1-x}Mn_xAs$ ($t = 10$ nm)/MnO($t \sim 4$ nm) bilayer measured at $T = 10$ K, after cooling in the presence of different magnetic fields: (a) $H = 2500$ Oe, (b) $H = -2500$ Oe and (c) $H = 0$. In (d), we show the hysteresis loop measured at $T = 10$ K for an uncapped $Ga_{1-x}Mn_xAs$ ($t = 15$ nm) control sample, after field cooling in $H = 1000$ Oe. The diamagnetic and/or paramagnetic background has been subtracted from these (and all subsequently presented) hysteresis loops.

**Figure 2.** Panels (a) – (c) provide a summary of the temperature dependence of the hysteresis loops and the hysteresis loop parameters for the sample described in Fig. 1. The hysteresis loop at any given temperature is measured after cooling the sample from $T = 100$ K to the measurement temperature in a field of $H = 2500$ Oe. Panels (a) and (b) show the field cooled hysteresis loop at $T = 5$ K and $T = 30$ K, respectively. Panel (c) shows the temperature dependence of the exchange field, $H_E = |(H_{C-}+H_{C+})/2|$, and the coercive field, $H_C = |(H_{C-}-H_{C+})/2|$ for $T < T_N$. The inset panel (d) shows the temperature dependent magnetization at $H = 20$ Oe, after field cooling in $H = 10$ kOe.

**Figure 3.** Hysteresis loop parameters at $T = 10$ K, as a function of the cooling field for the sample described in Fig. 1. Measurements are taken after cooling from $T = 100$ K to $T = 10$ K in the respective magnetic fields. Note that the field axis is split at $H = 1.1$ kOe and is depicted on different scales for magnetic fields less than or greater than this value.

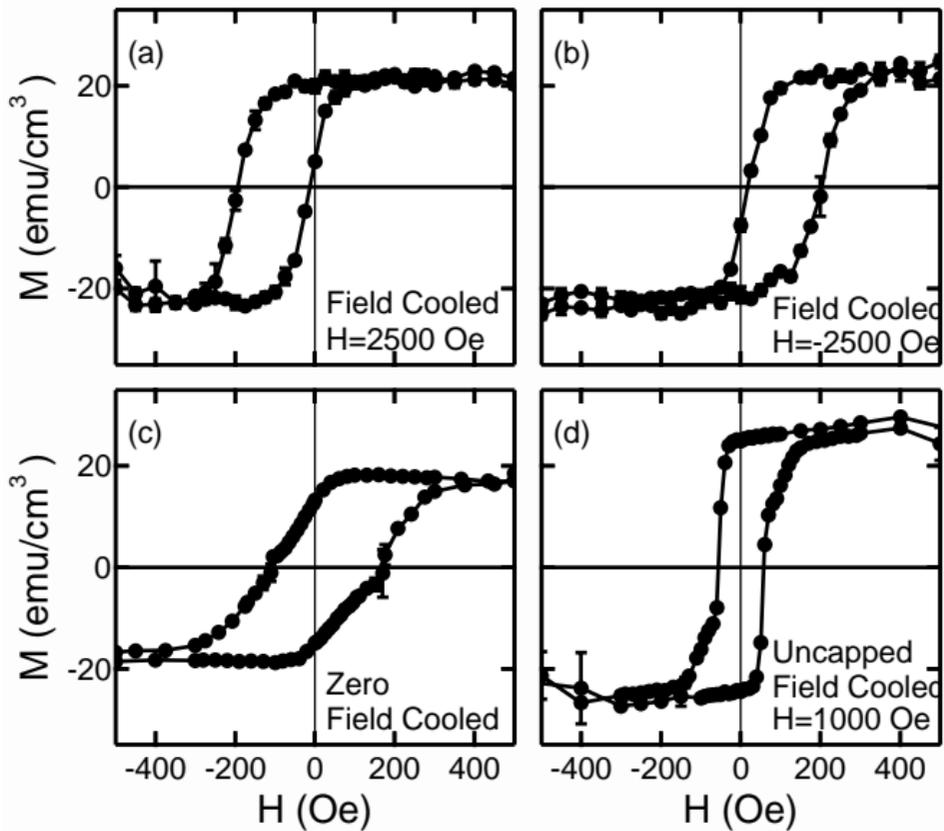

Figure 1, K. F. Eid *et al.*

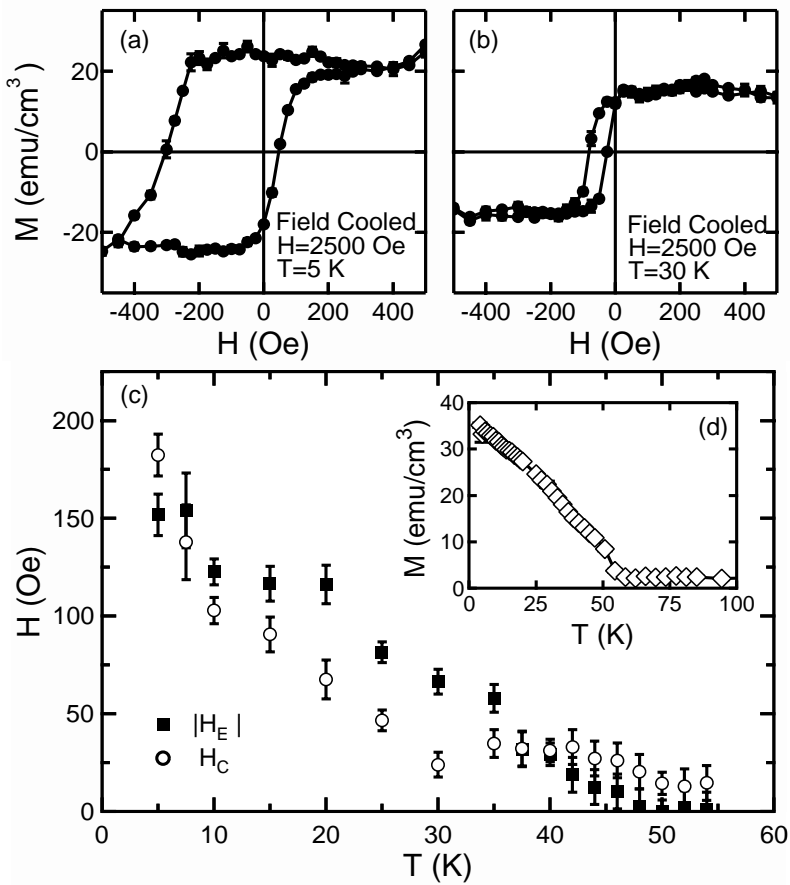

Figure 2, K. F. Eid *et al.*

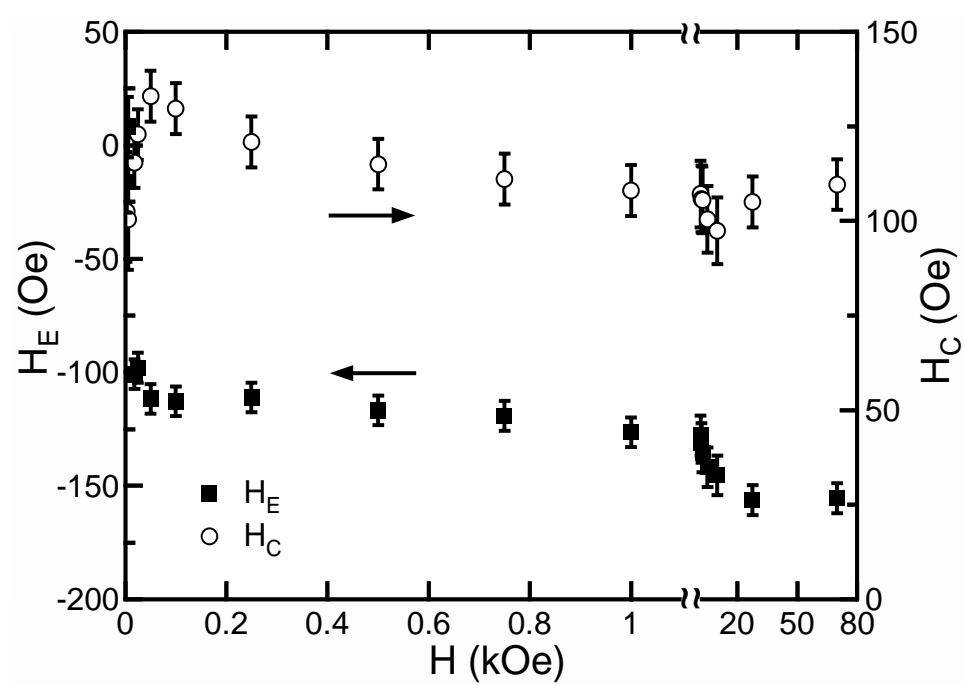

Figure 3, K. F. Eid *et al.*